\documentclass[sigconf]{acmart}

\AtBeginDocument{%
  }

\copyrightyear{2024}
\acmYear{2024}
\setcopyright{acmlicensed}\acmConference[MuC '24]{Proceedings of Mensch und Computer 2024}{September 1--4, 2024}{Karlsruhe, Germany}
\acmBooktitle{Proceedings of Mensch und Computer 2024 (MuC '24), September 1--4, 2024, Karlsruhe, Germany}
\acmDOI{10.1145/3670653.3677495}
\acmISBN{979-8-4007-0998-2/24/09}

\usepackage{amsmath} 
\usepackage{enumitem}
\newcommand{\ratingscale}{
  \begin{tabular}{@{}*{5}{c@{\hskip 0.5em}}}
  Strongly Disagree & Disagree & Neutral & Agree & Strongly Agree
  \end{tabular}
}

\begin{document}

\title{Design Frictions on Social Media: Balancing Reduced Mindless Scrolling and User Satisfaction}

\author{Nicolas Ruiz}
\affiliation{%
  \institution{University of Bremen}
  \city{Bremen}
  \country{Germany}}
\email{nruiz@uni-bremen.de}

\author{Gabriela Molina León}
\affiliation{%
  \institution{University of Bremen}
  \city{Bremen}
  \country{Germany}}
\email{molina@uni-bremen.de}

\author{Hendrik Heuer}
\affiliation{%
  \institution{Center for Advanced Internet Studies \& University of Wuppertal}
  \city{Bochum}
  \country{Germany}}
\email{hendrik.heuer@cais-research.de}

\renewcommand{\shortauthors}{Ruiz et al.}

\begin{abstract}
Design features of social media platforms, such as infinite scroll, increase users' likelihood of experiencing normative dissociation --- a mental state of absorption that diminishes self-awareness and disrupts memory. This paper investigates how adding design frictions into the interface of a social media platform reduce mindless scrolling and user satisfaction. We conducted a study with 30 participants and compared their memory recognition of posts in two scenarios: one where participants had to react to each post to access further content and another using an infinite scroll design. Participants who used the design frictions interface 
exhibited significantly better content recall, although a majority of participants found the interface frustrating. We discuss design recommendations and scenarios where adding design frictions to social media platforms can be beneficial.
\end{abstract}




\maketitle

\section{Introduction}
Social media platforms have a direct impact on users' well-being and mental health \cite{ sadagheyani_investigating_2020, berryman_social_2018, abi-jaoude_smartphones_2020}. The use of these platforms can increase feelings of depression and loneliness \cite{sadagheyani_investigating_2020}, especially among young people \cite{berryman_social_2018, abi-jaoude_smartphones_2020}. Researchers have suggested that these adverse effects of social media are not coincidental, but rather strategically designed \cite{monge_roffarello_defining_2023, monge_roffarello_towards_2022, bhargava_ethics_2021}. Features of social media known as \textit{attention-capture dark patterns}~\cite{monge_roffarello_defining_2023, monge_roffarello_towards_2022} are purposely integrated to capture user attention and increase time spent online. One design
feature commonly found across platforms is infinite scroll \cite{monge_roffarello_defining_2023}, where ``users trigger the dynamically loading of additional content by scrolling down the page'' \cite{rixen_loop_2023}. Despite its popularity and ease of use, infinite scroll is associated with mindless scrolling  \cite{mildner_ethical_2021} as it keeps users engaged in prolonged sessions \cite{monge_roffarello_defining_2023}, often leading to subsequent feelings of regret \cite{rixen_loop_2023}. Baughan et al.~\cite{baughan_i_2022} showed how the infinite feed of X can increase the propensity to experience normative dissociation ---an absorbed state of mind characterized by a loss of awareness \cite{butler_normative_2006} and a disruption in memory \cite{freyd_cognitive_1998}, rendering users unable to remember what they read online. To mitigate the adverse effects of attention-capture dark patterns, researchers and designers have developed external supports to help users self-limit and self-monitor their social media use. These tools operate on top of social media apps without changing their inner workings and can include timers and locks that limit the access to platforms \cite{zhang_monitoring_2022}. However, external supports are not application-specific and indiscriminately block access to content that people might still want to engage with \cite{lukoff_switchtube_2023, tran_modeling_2019}. An emerging and relatively unexplored approach involves redesigning the interfaces to eliminate internal mechanisms that are commonly perceived to negatively affect users' well-being \cite{zhang_monitoring_2022}. For instance, researchers have demonstrated that nudging users about content they have already seen helps them avoid scrolling through old posts and increases the session quality \cite{zhang_monitoring_2022}. Similarly, hiding distracting recommendations when using YouTube for educational purposes has shown positive effects on users' sense of agency \cite{lukoff_switchtube_2023}. Redesigning interfaces appears to be more effective in enhancing a sense of agency and well-being than external supports to self-limit and self-control social media usage~\cite{zhang_monitoring_2022}. 

Our paper contributes to this line of work by exploring how adding design frictions (i.e., points of difficulty in users' experience with technology) affects user satisfaction and mindless scrolling on social media. We drew inspiration from the notion of \textit{microboundaries} --- small moments of friction that are designed to interrupt automatic, mindless interactions by providing opportunities for reflection \cite{cox_design_2016}. We conducted a study to analyze how adding a \textit{microboundary} to users' interactions with social media posts affects mindless scrolling. Specifically, we made reacting to posts a prerequisite for accessing further content (reaction-based interface) and compared it with an infinite-scroll version of the same application. Following the normative dissociation framework, we operationalize mindless scrolling as a state of ``absorption and a diminished self-awareness, often accompanied by a reduced sense of time, control and a gap in one’s memory.'' \cite{baughan_i_2022} Thus, we assessed mindless scrolling by measuring participants' recognition memory of posts while interacting with both interfaces. Then, we measured user satisfaction when using the reaction-based interface, given that feelings of frustration are common when frictions are added to users' primary tasks in social media \cite{wang_field_2014, monge_roffarello_nudging_2023}. We found that participants who interacted with the reaction-based interface remembered the content of posts more often than those who used the interface featuring infinite scrolling. This suggests that our intervention can prevent users from experiencing a symptom of normative dissociation, namely, a disruption in the ability to recall information. However, more than half of the participants (53\%) felt frustrated by having to react to each post, and only three out of 15 (20\%) indicated that they would like to use platforms that include this feature. We discuss design implications, such as taking situational needs into account and giving users the option to switch back to an interface with infinite scroll to reduce feelings of frustration. We also discuss scenarios where making users react to content can be useful, such as educational apps that require a more present state of mind.

\section{Related Work}
Infinite scroll is a design functionality present in almost all social media platforms that automatically adds new content to the bottom of the page as people scroll down their feeds \cite{rixen_loop_2023}. Despite its advantages, studies indicate that infinite scroll is not a neutral design pattern, but rather one of the main causes why people mindlessly scroll on social media \cite{mildner_ethical_2021}. Infinite scroll promotes endless sessions by reducing people's mental and physical effort \cite{widdicks_backfiring_2020, monge_roffarello_defining_2023}. To prevent mindless scrolling, researchers have suggested incorporating custom lists into the interface of Twitter to nudge users when they exhausted new content \cite{baughan_i_2022}. This intervention helped people feel more in control of their scrolling compared to when the content was shown all together in the defaulted infinite feed. Similarly, Monge and De Russis \cite{monge_roffarello_nudging_2023} found that nudging users when scrolling too quickly increased participants' feelings of control over their social media use. 

Infinite feeds have also been associated with an increased likelihood of people experiencing normative dissociation symptoms \cite{baughan_i_2022}. According to Butler \cite{butler_normative_2006}, states of normative dissociation are characterized by a high level of absorption where people experience low levels of self-awareness, reflective consciousness, and intention. One of the most common characteristics is a disruption in memory known as \textit{highway hypnosis}. Freyd et al. \cite{freyd_cognitive_1998} describe this experience as the loss of awareness while performing an activity, such as driving a car, and then switching back to a conscious state of mind without a clear memory of what happened. However, it is important to note that these experiences are natural and common cognitive processes that people engage with when performing activities like reading a book, exercising, or walking \cite{butler_normative_2006}. States of \textit{flow} \cite{snyder_handbook_2001}, for example, where there is a dynamic interplay between challenges and skills as people engage in meaningful activities, are highly rewarding and tend to encourage repetition. Normative dissociation experiences on social media become problematic when it reduces the volition and sense of control of the users, making them feel that their objectives regarding social media use have not been met \cite{baughan_i_2022}. Studying mindless scrolling through the normative dissociation framework does not only provide us a quantitative assessment method, but also a better understanding that takes into account the expectations, intentions, and goals of the users.  

To avoid automatic and thoughtless behaviors while interacting with technology, researchers have begun advocating for the integration of design frictions into user interfaces \cite{cox_design_2016, gould_special_2021, mejtoft_design_2023, mejtoft_design_2019}. Design frictions are defined as points of difficulty encountered during the interaction with technology \cite{cox_design_2016}, typically to prevent mistakes (e.g., a pop-up message to confirm an action). In UX/UI design, the general belief is that the less friction a platform has, the more seamless, effortless, and painless the interactions will be, and therefore, frictions should be avoided at all costs. However, researchers have started to explore the benefits of purposely adding design frictions into peoples' interaction with technology to promote more mindful experiences \cite{cox_design_2016}. Wang et al. \cite{wang_field_2014}, for instance, showed adding time delays before posting increased self-reflection among Facebook users and helped them avoid heated online discussions. Similarly, Haliburton et al. \cite{haliburton_longitudinal_2024} showed that imposing customizable short delays when users opened a target app increased intentional use over time. Lyngs et al. demonstrated that goal reminders helped users to stay on task while using Facebook, although with the risk of the intervention becoming annoying \cite{lyngs_i_2020}. Our study builds on this line of work by exploring the effect of adding small frictions to users' interaction with a social media feed to prevent mindless scrolling. 

\section{Methodology}
We conducted a between-subjects study to investigate the effect of design frictions on mindless scrolling and satisfaction. Since design frictions can be wide-ranging and encompass different types of interventions \cite{cox_design_2016}, we opted to design a reaction-based interface of a social media application that makes reacting to posts a prerequisite to engage with further content and compared it with an infinite-scroll version of the same application. Therefore, we ask (1) \textit{How does the reaction-based interface affect mindless scrolling compared to the infinite-scroll interface?} and (2) \textit{How does the reaction-based interface affect user experience?} We operationalize mindless scrolling using the normative dissociation framework and measured the ability of the participants to recall content through an old-new memory recognition test \cite{zimmerman_foodie_2020}. Additionally, we elicited participant impressions of the interface containing design frictions through a survey. 

\begin{figure*}
    \centering
    \includegraphics[width=\textwidth]{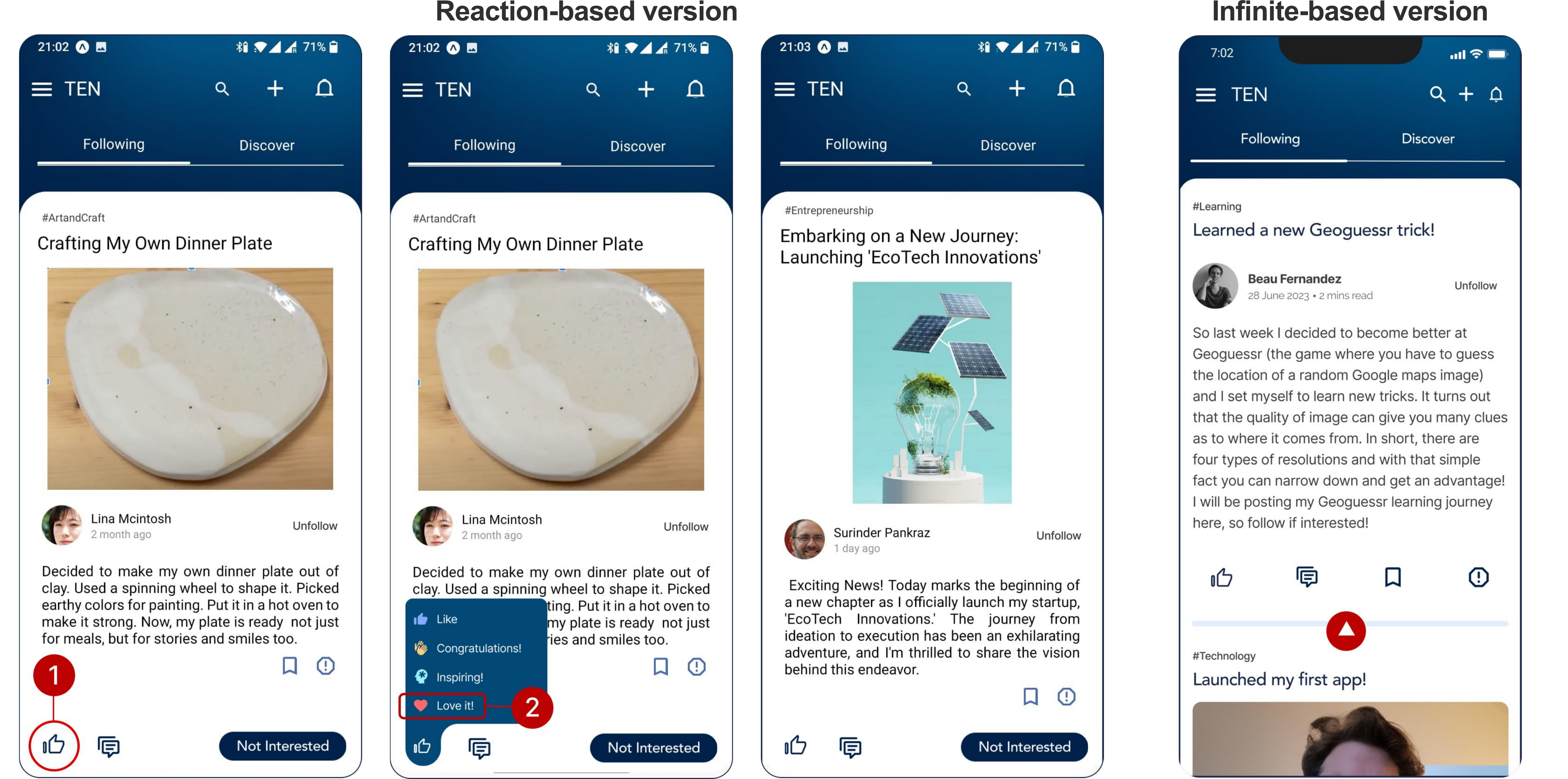}
    \Description{This figure displays three mobile phone screens, each containing a header and a feed of posts. The first two screens belong to the reaction-base interface while the last one to the infinite-scroll interface The first screen illustrates a user pressing a "like" button, which reveals four reaction options above it. The second screen shows a different post, indicating that the user has selected one of the reaction options and moved to the next content. The third screen demonstrates the infinite scroll feature, displaying two posts stacked vertically to represent the continuous feed of the app.}
    \caption{In the \textit{reaction-based version}, participants have to either select one of four reactions on the left or the \textit{not interested} button to see the next post. In the \textit{infinite-scroll version}, participants can scroll up and down without restrictions.}
    \label{fig:phones}
\end{figure*}

\subsection{Reaction-Based and Infinite-scroll Interfaces}
Both reaction-based (intervention) and infinite-scroll (control condition) interfaces were designed for a fictitious social media application, each containing user-generated posts displayed in a feed, as shown in Figure \ref{fig:phones}. The only difference between the conditions was the mechanism users employed to interact with content; other elements of the feed, such as content, pictures, users, and the order of posts, remained unchanged. In the infinite-scroll version, the content was rendered dynamically as the person scrolled down the feed. In contrast, the reaction-based version required participants to react to the current post to view the next one. If participants did not wish to react to a post, they had the option to press a ``not interested'' button to skip to the next one.

Both versions of the application contained the same image and text-based posts, similar to those on platforms like Facebook and Reddit. As the ability to recall content can be highly sensitive to the topic of the posts, we included 10 different categories: \textit{Art, Cooking, Learning, Sports, Personal development, Entrepreneurship, Technology, Yoga, Research}, and \textit{Hobbies}. We extracted and modified the post texts from \textit{subreddits} of Reddit dedicated to those categories (e.g., \textit{r/painting} and \textit{r/cooking}). The subreddits for each category were selected based on their popularity, measured by the number of members. The criteria for selecting the posts were the quality and size of the text. Due to privacy reasons, we used images from free stock image websites. Moreover, each post included a reaction button. This button was modeled after the popular reactions available on Facebook. and displayed four different reactions for participants to select: ``Like'', ``Congratulations!'', ``Inspiring!'', and ``Love it!''. These reactions were selected for their positive and encouraging tone, as the posts focused on people's progress in personal projects and hobbies. Offering four reaction options instead of a "next post" button was intended to encourage participants to pause briefly and select the reaction that best aligned with their perception of the content.
\subsection{Participants}
The sample consisted of 30 university students recruited using snowball sampling through the university's mailing list and direct messages to personal contacts. The study was also promoted via posters placed around the university campus, indicating that we aimed to investigate "mindful social media use and digital well-being." The age range spanned from 21 to 31 (std = 2.603), as this group falls within the age range that most frequently uses social media platforms like Instagram \cite{i_wearesocial2024} and Facebook \cite{f_wearesocial2024}. Twenty participants identified as male and 10 as female. 

\subsection{Procedure}
 First, we asked participants to give their informed consent and fill out a demographics form in paper format. Then, they completed a survey to elicit their social media behavior and previous experiences of normative dissociation on social media. Upon completion, participants underwent two phases of an old-new recognition test --- an exposure phase and a test phase \cite{zimmerman_foodie_2020}. The primary objective of the test was to draw conclusions about memory performance by calculating the memory sensitivity of each participant (d'), which assessed their ability to accurately recognize stimuli they had previously encountered as ``old'' and distinguish them from those they had not seen before as ``new'' \cite{toth_eeg_2021}. During the exposure phase, participants used a smartphone with the app preinstalled. In the pilot study, 50 posts were perceived as overwhelming, so we decided to include 30 posts. We asked participants to browse the feed as they would in any social media app. They were not informed that their capacity to remember the content of the posts would be tested later. Then, participants performed the Stroop task to reduce memory performance, as retention of images tends to be high in the short term \cite{zimmerman_foodie_2020}. In this task, participants were presented with color words written in mismatched colors (e.g., the word ``yellow'' written in red) and were asked to name the color of the word. Next, we tested their memory recognition ability by giving them a mobile phone with 30 posts, out of which 20 had been shown to them in the exposure phase, and 10 were new. Each post included two buttons: \textit{new} and \textit{old}. Participants had to select one to indicate whether they remembered seeing the post beforehand. At the end, the participants from the treatment group responded to a five-point Likert-scale questionnaire to collect their impressions of the reaction-based interface and their feelings of frustration, as one of the main challenges of designing frictions is determining the right amount of friction that does not interfere with a seamless user experience (see questionnaire in the \hyperref[appendix]{appendix}). Participants in the intervention group took circa 20 minutes to complete the study, while those in the control group took 15 minutes.

\section{Results}
According to the survey responses, more than half of the participants either agreed or strongly agreed that they spent more time on social media than initially intended and that they lost track of time while doing so (53\%). Moreover, only half of the participants indicated that they remember most of the content they consume on social media. Twenty-one participants (70\%) thought it was important to remember the content they engage with. This suggests that many participants acknowledged experiencing symptoms of normative dissociation while using social media, specifically a distortion in their perception of time, leading to long sessions and a decrease in the ability to recall content.

A Mann-Whitney U test on the old-new recognition memory test scores indicated a significant difference in memory sensitivity (d') between the control and treatment groups ($U = 31, p < 0.001$). Memory sensitivity (d') is widely used in old-new recognition tests and calculated using the \textit{z transforms} of the hit rate \textit{H} (correctly identified "old" stimuli) and the false alarm rate \textit{F} (incorrectly identified "new" stimuli) \cite{hautus_detection_2021}. Participants of the treatment group demonstrated a superior self-reported memory recognition (d') of the content they engaged with during the exposure phase ($m = 2.89$), compared to those who interacted with the infinite-scroll version ($m = 1.20$). A rank-biserial correlation indicated a large effect size of the intervention (\( R_{\text{rb}} = 0.72 \)). This suggests that requiring reactions to each post can be effective in preventing one symptom of normative dissociation --- the self-reported reduced memory of the experience. Participants took significantly longer to interact with the reaction-based interface than with the infinite-scroll version ($U = 9, p < 0.001$), which might help explain the difference in recall performance. On average, the intervention group took 8.67 (std = 2.690) minutes to scroll though all posts during the exposure phase, whereas the control group took only 3.33 (std = 1.589).

Furthermore, the effect of the reaction-based interface was also perceived by the participants --- 67\% either strongly agreed or agreed that this interface was effective in making them pay more attention to the content of the posts. However, more than half of the participants felt frustrated about having to react to each post (53\%) and one in three (33\%) indicated that they felt demotivated to continue using the app because of this feature. After concluding the experiment, a participant commented that the reason why he disliked the reactions is that he identified as a ``lurker'', and thus, prefers to remain unseen on social media platforms. Moreover, only 20\% of the participants indicated that they would like to see more applications with this feature in the future. This shows a misalignment between the perceived usefulness of attentiveness while using social media and the willingness of the participants to engage with the proposed interface, which was mostly considered to be effective in increasing attention and memory of content. 

\section{Discussion}
Our findings align with the observations of Wang et al. \cite{wang_field_2014} regarding the inclusion of time delay before publishing social media posts --- adding extra steps in interactions is regarded as both beneficial and annoying. Although we anticipated this effect and included a "not interested" button for skipping posts without interacting, it was insufficient to prevent frustration. Based on our findings, we discuss recommendations for integrating frictions into users' interactions with social media feeds.

First, to reduce the frustration likelihood while still leveraging the benefits of the reaction-based interface, we encourage designers and developers to create a less intrusive intervention that prompts users to react to posts at regular intervals (e.g., every 5-10 posts) instead of every time. This approach might be seen as less intrusive while still increasing attention. Moreover, the frequency in which users are required to react could also be left to individual preference. As Haliburton et al. \cite{haliburton_longitudinal_2024} showed, users benefit from customizing the duration of time delays while opening target apps based on personal requirements. Future work can explore the effect of varying the reaction frequency on user experience.
Second, there are scenarios where a reaction-based interface  might still be beneficial, especially when a high level of involvement is desired. As highlighted by Lukoff et al. \cite{lukoff_how_2021}, goal-oriented use that satisfies informational needs (e.g., searching for tutorials on YouTube) can be supported by more restrictive interfaces, which enhance users' sense of agency. Encouraging people to engage with posts on goal-oriented sessions could improve information recall and attention. Future research could investigate the impact of reaction-based interfaces on user experience on platforms such as LinkedIn, YouTube, or Medium, where informational and instrumental usage tends to be more prevalent.
Lastly, social media platforms could also benefit from incorporating a reaction-based interface to assist users in managing their social media use. 
For instance, providing users the option to switch between a less permissive interface of YouTube without distracting recommendations and its standard version, depending on the intended use, results in increased satisfaction and alignment with goals \cite{lukoff_switchtube_2023}. Similarly, experiences of normative dissociation are generally perceived negatively when users feel they have wasted time and cannot recall what they read. However, becoming absorbed while browsing can also be considered a positive and beneficial experience by offering relief and escape from the present moment \cite{baughan_i_2022}.Therefore, considering situational needs and contextual use is vital for the design of positive experiences online.  We could offer a reaction-based interface for users who wish to regulate their social media use and avoid mindless scrolling, and an infinite scroll interface for when browsing in autopilot is not a concern.

Our study has of course limitations. The generalizability of our findings is limited by a lack of gender diversity and the focus on a student sample. The ability to remember content on social media is not the only symptom of normative dissociation experiences. Future work could assess other symptoms such as loss of self-awareness and passage of time. Mindless scrolling is also influenced by users' interest and engagement with content, so the selected posts could have influenced these factors.  Although we included posts of different categories, it was still small compared to the content available on social media. Future research should explore the impact of a reaction-based interface within 

\section{Conclusion}
We showed how design frictions, through a reaction-based interface, can deter users from engaging in mindless scrolling on social media. Our results indicate that the reaction-based interface increased users' self-reported attention and memory recognition of posts compared to an infinite-scroll interface. This suggests the effectiveness of our intervention in preventing mindless scrolling, albeit at the cost of frustrating users. We suggested changing the reaction prompt frequency to reduce feelings of discomfort, as well as considering intentional needs. Moreover, we argued that educational-oriented apps can benefit from the reaction-based interface by increasing users' attention.

\begin{acks}
The authors would like to thank Hemanth Kumar, Disha Mukre, Priyanka Joshi, and Md Touhidul Islam for their work in designing and implementing the applications, as well as conducting the study.
\end{acks}

\bibliographystyle{ACM-Reference-Format}
\bibliography{main}

\appendix

\section*{General survey}
\label{appendix}
\begin{enumerate}[label=\arabic*.,leftmargin=*]
  \item I find myself spending more time on social media platforms than initially intended.
  
  \ratingscale
  
  \item I lose track of time when using social media platforms.
  
  \ratingscale 
  
  \item It is important to me to recall the content I find valuable on social media.
  
  \ratingscale 
  
  \item At the end of the day, I remember most of the content I consume on social media platforms.
  
  \ratingscale 
  \item I use social media without really paying attention to what I am doing.
  
  \ratingscale
  
\end{enumerate}

\section*{Reaction-based interface survey}

\begin{enumerate}[label=\arabic*.,leftmargin=*]
  \item I was motivated to react to each post.
  
  \ratingscale 
  
  \item It was frustrating having to react to every post before going to the next one.
  
  \ratingscale 
  \item I felt demotivated to continue using the app every time that I had to react to a post.
  
  \ratingscale 
  \item Having to react to each post made me pay more attention to the content of the posts.
  
  \ratingscale 
  \item I would like to see more social media platforms having this feature (i.e., having to react to each post).
  
  \ratingscale 
  \item I found the interface easy to interact with.
  
  \ratingscale 
  \item The screen layout was well organized.
  
  \ratingscale 
  \item Are there any changes or improvements you would recommend regarding the current interaction mechanism?
  
\end{enumerate}

\end{document}